\documentclass{webofc}

\usepackage[varg]{txfonts}   
\usepackage{hyperref}
\usepackage{url}
\usepackage{tikz-feynhand}
\usepackage{tikz-feynman}
\hypersetup{colorlinks=true,citecolor=blue,urlcolor=blue,linkcolor=blue}
\begin{document}
\begin{flushright}
OU-HET-1244
\end{flushright}

\title{CP violation of the loop induced $H^\pm \to W^\pm Z$ decays \\ in general two Higgs doublet model \thanks{Proceedings of the 2024 International Workshop on Future Linear Colliders (LCWS2024), Jul 8–11, 2024, The University of Tokyo, Japan} }

%
%

\author{\firstname{Shinya} \lastname{Kanemura}\inst{1} \fnsep \thanks{\email{kanemu@het.phys.sci.osaka-u.ac.jp}} \and
        \firstname{Yushi} \lastname{Mura}\inst{1}\fnsep \thanks{\email{y_mura@het.phys.sci.osaka-u.ac.jp}} \fnsep\thanks{Speaker}
}

\institute{Department of Physics, Osaka University, Toyonaka, Osaka 560-0043, Japan}

\abstract{
New sources of CP violation are necessary to solve the problem of the baryon asymmetry of the Universe.
Extending Higgs sector is one way to introduce such new CP violating phases, and studying observables resulting from the CP violation is important to test the model in future experiments.
In these proceedings, we discuss the loop induced $ H^\pm W^\mp Z$ vertices in the CP violating general two Higgs doublet model, summarizing our results~\cite{Kanemura:2024ezz}.
We evaluate impacts of the CP violation on the decays $H^\pm \to W^\pm Z$ through these vertices, and find that the difference between the decays $H^+ \to W^+ Z$ and $H^- \to W^- Z$ is sensitive to the CP violating phases in the model.
}
\maketitle
\section{Introduction}

By the discovery of the 125 GeV Higgs boson at the LHC in 2012~\cite{ATLAS:2012yve,CMS:2012qbp}, the Standard Model (SM) was established.
However, global structure of the Higgs potential is still unknown, and there are still some problems which cannot be solved by the SM.
The origin of the baryon asymmetry of the Universe is one of these problems.
Baryogenesis is a promising scenario to solve this problem, and the CP violation is required by Sakharov's third conditions~\cite{Sakharov:1967dj}.
However, the CP violation in the SM is not sufficient to explain the observed baryon asymmetry~\cite{Huet:1994jb} with the mechanism of electroweak baryogenesis~\cite{Kuzmin:1985mm}.
In the Two Higgs Doublet Model (2HDM), new sources of the CP violation can be introduced, so that electroweak baryogenesis in the 2HDM has been studied in many literature~\cite{Turok:1990zg,Cline:1995dg,Fromme:2006cm,Cline:2011mm,Tulin:2011wi,Liu:2011jh,Ahmadvand:2013sna,Chiang:2016vgf,Guo:2016ixx,Fuyuto:2017ewj,Dorsch:2016nrg,Modak:2018csw,Basler:2021kgq,Enomoto:2021dkl,Enomoto:2022rrl,Kanemura:2023juv,Aoki:2023xnn}.
Such new CP violating effects may appear in physical observables, e.g. electric dipole moment, so that it is important to study the CP violating observables for testing the model in future experiments.

In these proceedings, based on our results~\cite{Kanemura:2024ezz}, we discuss the CP violating effects of the loop induced $H^\pm W^\mp Z$ vertices in the general 2HDM.
The loop induced $H^\pm W^\mp Z$ vertices~\cite{Grifols:1980uq,Rizzo:1989ym} have been well studied as a consequence of the violation of the custodial symmetry, which is a global symmetry in the Higgs potential~\cite{Sikivie:1980hm,Haber:1992py,Pomarol:1993mu,Haber:2010bw,Gerard:2007kn,Grzadkowski:2010dj,Kanemura:2011sj,Aiko:2020atr}.
These vertices have been calculated at the one-loop level in the CP conserving 2HDM~\cite{Rizzo:1989ym,CapdequiPeyranere:1990qk,Kanemura:1997ej,Diaz-Cruz:2001thx,Arhrib:2006wd,Abbas:2018pfp,Aiko:2021can} and the Minimal Supersymmetric Standard Model (MSSM)~\cite{Mendez:1990epa,CapdequiPeyranere:1990qk,Arhrib:2007ed,Arhrib:2006wd,Arhrib:2007rm}.
In ref.~\cite{Kanemura:2024ezz}, the full formulae for these vertices in the general 2HDM have been calculated at the one-loop level.
It has been known that the CP violating Higgs potential also violates the custodial symmetry~\cite{Pomarol:1993mu}.
We discuss the effects to the $H^\pm W^\mp Z$ vertices, which are caused by the custodial symmetry violation through the CP violation in the general 2HDM.

\section{The custodial and CP symmetry in the general two Higgs doublet model}
\label{model}

We consider the most general 2HDM.
Two $\mathrm{SU}(2)_L$ doublets with the hypercharge $Y=1/2$ in the Higgs basis~\cite{Davidson:2005cw}
\begin{align}
  \Phi_1 = 
  \begin{pmatrix}
     G^+ \\
     \frac{1}{\sqrt{2}} ( v + h_1 + i G^0)
  \end{pmatrix},
  ~~
  \Phi_2 = 
  \begin{pmatrix}
     H^+ \\
     \frac{1}{\sqrt{2}} ( h_2 + i h_3)
  \end{pmatrix},
\end{align}
are introduced, where $v$ ($\simeq 246$ GeV) is the Vacuum Expectation Value (VEV), $G^+$ and $G^0$ are the Nambu--Goldstone bosons, $H^\pm$ and $h_i$ ($i=1,2,3$) are the charged and the neutral scalar bosons, respectively.
The scalar potential is given by
\begin{align}
  V = &-Y_1^2 (\Phi_1^\dagger \Phi_1) - Y_2^2 (\Phi_2^\dagger \Phi_2) - \Big(Y_3^2 (\Phi_1^\dagger \Phi_2) + \mathrm{h.c.} \Big) \notag \\
  &+\frac{1}{2}Z_1 (\Phi_1^\dagger \Phi_1)^2 +\frac{1}{2}Z_2 (\Phi_2^\dagger \Phi_2)^2 + Z_3 (\Phi_1^\dagger \Phi_1)(\Phi_2^\dagger \Phi_2) + Z_4 (\Phi_2^\dagger \Phi_1)(\Phi_1^\dagger \Phi_2) \notag \\
  &+ \bigg\{ \Big( \frac{1}{2}Z_5 \Phi_1^\dagger \Phi_2 + Z_6 \Phi_1^\dagger \Phi_1 + Z_7 \Phi_2^\dagger \Phi_2 \Big) \Phi_1^\dagger \Phi_2 + \mathrm{h.c.} \bigg\},
  \label{eq:potential}
\end{align}
where $Y_1^2 (>0), Y_2^2, Z_1, Z_2, Z_3, Z_4 \in \mathbb{R}$ and $Y_3^2, Z_5, Z_6, Z_7 \in \mathbb{C}$.
By using a degree of freedom of rephasing for $\Phi_2$, we can make $Z_5$ real.
The stationary conditions are given by
\begin{align}
  Y_1^2 = \frac{1}{2} Z_1 v^2, ~~~~ Y_3^2 = \frac{1}{2} Z_6 v^2.
\end{align}
The squared mass of $H^\pm$ and the squared mass matrix $\mathcal{M}_{ij}^2 \equiv \partial^2 V / \partial h_i \partial h_j $ for neutral scalar bosons are given by
\begin{align}
  &m_{H^\pm}^2 = -Y_2^2 + \frac{1}{2} Z_3 v^2,
  \notag \\
  &\mathcal{M}_{ij}^2 = 
  \begin{pmatrix}
      Z_1 v^2 & Z_6^R v^2 & - Z_6^I v^2 \\
      Z_6^R v^2 & m_{H^\pm}^2 +\frac{1}{2}(Z_4 + Z_5) v^2 & 0 \\
      - Z_6^I v^2 & 0 & m_{H^\pm}^2 +\frac{1}{2}(Z_4 - Z_5) v^2
  \end{pmatrix}.
  \label{eq:masses}
\end{align}
We define the mass eigenstates for the neutral scalar bosons $H_i$ as 
\begin{align}
  H_i = \mathcal{R}_{ij} h_j,
  \label{eq:definemasseigen}
\end{align}
where $\mathcal{R} \in \mathrm{SO}(3)$ satisfies 
\begin{align}
  \mathcal{R} \mathcal{M}^2 \mathcal{R}^T = \mathrm{diag}(m_{H_1}^2, m_{H_2}^2, m_{H_3}^2).
  \label{eq:relation_neutmass_diagmass}
\end{align}
The orthogonal matrix $\mathcal{R}$ is parametrized as 
\begin{align}
  \mathcal{R} =
  \begin{pmatrix}
      \cos \alpha_2 \cos \alpha_1 & - \cos \alpha_3 \sin \alpha_1 - \cos \alpha_1 \sin \alpha_2 \sin \alpha_3 & -\cos \alpha_1 \cos \alpha_3 \sin \alpha_2 + \sin \alpha_1 \sin \alpha_3 \\
      \cos \alpha_2 \sin \alpha_1 & \cos \alpha_1 \cos \alpha_3 - \sin \alpha_1 \sin \alpha_2 \sin \alpha_3 & -\cos \alpha_3 \sin \alpha_1 \sin \alpha_2 - \cos \alpha_1 \sin \alpha_3 \\
      \sin \alpha_2 & \cos \alpha_2 \sin \alpha_3 & \cos \alpha_2 \cos \alpha_3
  \end{pmatrix},
  \label{eq:Rmatrix}
\end{align}
where $-\pi \le \alpha_1, \alpha_3 < \pi$, $-\pi/2 \le \alpha_2 < \pi/2$~\cite{Haber:2006ue}.
We set $m_{H_1} = 125$ GeV to identify $H_1$ with the SM Higgs boson.

The Yukawa interaction is given by
\begin{align}
  -\mathcal{L}_Y &=  \overline{Q_{L}^u} Y^u_{\mathrm{d}} \tilde{\Phi}_1 u_{R} + \overline{Q_{L}^d} Y^{d}_{\mathrm{d}} \Phi_1 d_{R} + \overline{L_{L}} Y^e_{\mathrm{d}} \Phi_1 e_{R} + \mathrm{h.c.}  \notag \\
  &+ \overline{Q_{L}^u} \rho^u \tilde{\Phi}_2 u_{R} + \overline{Q_{L}^d} \rho^{d} \Phi_2 d_{R} + \overline{L_{L}} \rho^e \Phi_2 e_{R} + \mathrm{h.c.},
\end{align}
where the $\mathrm{SU}(2)_L$ quark and lepton doublets are defined by $Q_L^u =  (u_L, V_{\mathrm{CKM}} d_L)^T$, $Q_L^d =  ( V_{\mathrm{CKM}}^\dagger u_L,d_L)^T$ and $L_L = (\nu_L, e_L)^T$.
The Yukawa matrices $Y_{\mathrm{d}}^f$ ($f = u,d,e$) among the SM fermions and $\Phi_1$ are written by $Y_{\mathrm{d}}^f = \frac{\sqrt{2}}{v} \mathrm{diag}(m_{f_1}, m_{f_2}, m_{f_3})$.
On the other hand, in general, $\rho^f$ matrices are not diagonalized.
We parameterize these matrices as 
\begin{align}
  \rho^u = 
  \begin{pmatrix}
  \rho_{uu} & \rho_{uc} & \rho_{ut} \\
  \rho_{cu} & \rho_{cc} & \rho_{ct} \\
  \rho_{tu} & \rho_{tc} & \rho_{tt}
  \end{pmatrix}, 
  ~~
  \rho^d = 
  \begin{pmatrix}
  \rho_{dd} & \rho_{ds} & \rho_{db} \\
  \rho_{sd} & \rho_{ss} & \rho_{sb} \\
  \rho_{bd} & \rho_{bs} & \rho_{bb}
  \end{pmatrix},
  ~~
  \rho^e = 
  \begin{pmatrix}
  \rho_{ee} & \rho_{e\mu} & \rho_{e\tau} \\
  \rho_{\mu e} & \rho_{\mu \mu} & \rho_{\mu \tau} \\
  \rho_{\tau e} & \rho_{\tau \mu} & \rho_{\tau \tau}
  \end{pmatrix}.
\end{align}
The components of each matrix are generally complex, so that the CP violating phases can be introduced.

We then discuss the custodial symmetry and the CP symmetry in our model.
According to ref.~\cite{Haber:2010bw}, we introduce two bilinear forms as
\begin{align}
  \mathbb{M}_1 \equiv \big(\tilde{\Phi}_1, \Phi_1 \big), ~~  \mathbb{M}_2 \equiv \big(\tilde{\Phi}_2, \Phi_2 \big)
  \begin{pmatrix}
    e^{- i\chi}  & 0 \\
    0 & e^{i\chi}
  \end{pmatrix},
\end{align}
where the phase $\chi$ ($0 \le \chi < 2\pi$) represents the degree of freedom of the phase rotation for $\Phi_2$.
The transformation law of $\mathbb{M}_{1,2}$ under global $\mathrm{SU}(2)_L \times \mathrm{SU}(2)_R$ is $\mathbb{M}_{1,2} \to L \mathbb{M}_{1,2} R^\dagger$, where $L \in \mathrm{SU}(2)_L$ and $R \in \mathrm{SU}(2)_R$.
If the scalar potential given in eq.~(\ref{eq:potential}) is invariant under the global $\mathrm{SU}(2)_L \times \mathrm{SU}(2)_R$ transformation, it is invariant under the $L=R$ transformation even after the spontaneous electroweak symmetry breaking.
This remaining symmetry is so-called the custodial $\mathrm{SU}(2)_V$ symmetry.

The conditions for the custodial symmetric scalar potential are given by~\cite{Pomarol:1993mu,Gerard:2007kn,Haber:2010bw}
\begin{align}
  &\mathrm{Im}[Y_3^2 e^{- i \chi}] = \mathrm{Im}[Z_5 e^{-2 i \chi}] =  \mathrm{Im}[Z_6 e^{- i \chi}] = \mathrm{Im}[Z_7 e^{- i \chi}] = 0, \notag \\
  &Z_4 = \mathrm{Re}[Z_5 e^{-2 i \chi}].
\end{align}
If we take the basis of $\Phi_2$ such that $Z_5$ is real, the conditions are 
\begin{align}
  &\mathrm{Custodial~symmetry}: \hspace{1.65cm} Z_4 = Z_5, ~~~~ Z_6^I = Z_7^I = 0 ~~~~ (\chi = 0, ~\pi),
  \label{eq:condition_custodial} \\
  &\mathrm{Twisted~custodial~symmetry}:~~Z_4 = -Z_5, ~~ Z_6^R = Z_7^R = 0 ~~~ (\chi = \pi/2, ~3 \pi/2).
  \label{eq:condition_twist_custodial}
\end{align}
Eqs.~(\ref{eq:condition_custodial}) and (\ref{eq:condition_twist_custodial}) have been known as the conditions for the custodial and twisted custodial symmetry in the potential, respectively~\cite{Pomarol:1993mu,Gerard:2007kn,Haber:2010bw}.
We note that even if the potential is custodial symmetric, in the gauge and Yukawa sector, the custodial symmetry is violated.

On the other hand, the conditions for the CP symmetric scalar potential are given by~\cite{Botella:1994cs,Davidson:2005cw,Gunion:2005ja,Haber:2010bw}
\begin{align}
  Z_5 \mathrm{Im}[Z_6^2] = Z_5 \mathrm{Im}[Z_7^2] = \mathrm{Im}[Z_6^* Z_7] = 0.
  \label{eq:condition_CPV} 
\end{align}
in the real $Z_5$ basis.
When any of these rephasing invariants is non-zero, the potential violates the CP symmetry.
Therefore, by comparing eqs.~(\ref{eq:condition_custodial})-(\ref{eq:condition_CPV}), it can be understood that the custodial symmetry is violated by the CP violating potential.

The additional Yukawa matrices $\rho^f$ $(f=u,d,e)$ in our model are also able to violate the CP symmetry.
For example, when we consider the case of $\rho^f = 0$ except for $\rho_{tt}$, in addition to the quantities shown in eq.~(\ref{eq:condition_CPV}), all of $Z_5 \mathrm{Im}[\rho_{tt}^2]$, $\mathrm{Im}[Z_6 \rho_{tt}]$ and $\mathrm{Im}[Z_7 \rho_{tt}]$ have to be zero for the CP symmetry.

In the next section, we analyze the loop induced $H^\pm W^\mp Z$ vertices as a consequence of the custodial symmetry violation.
We also discuss the CP violating effects to these vertices.

\section{The $H^\pm \to W^\pm Z$ decays and the CP violation}
In this model, the $H^\pm W^\mp Z$ vertices induced at the loop level.
We define these vertices as
\begin{align}
    m_W g V_{\mu \nu}^{\pm} = 
    \begin{minipage}[b]{0.25\linewidth}
    \centering
    \setlength{\feynhandlinesize}{0.7pt}
    \setlength{\feynhandarrowsize}{4pt}
    \begin{tikzpicture}[baseline=-0.15cm]
    \begin{feynhand}
    \vertex (a) at (0,0){$H^\pm$}; \vertex[NWblob] (d) at (1.5,0) {};
    \vertex (e) at (2.2,-1.3){$Z_\nu$};\vertex (f) at (2.2,1.3){$W^\pm_\mu$};
    \propag[chasca, mom={$k_1$}] (a) to (d);
    \propag[bos, mom={$k_3$}] (d) to (e);
    \propag[chabos, mom={$k_2$}] (d) to (f);
    \end{feynhand}
    \end{tikzpicture}
    \end{minipage}.
\end{align}
The tensor $V_{\mu \nu}^{\pm}$ are decomposed by
\begin{align}
  V_{\mu\nu}^{\pm} = F_{\pm} g_{\mu \nu} + \frac{G_{\pm}}{m_W^2} k^3_\mu k^2_\nu  + \frac{H_{\pm}}{m_W^2} \epsilon_{\mu \nu \rho \sigma}k_3^\rho k_2^\sigma.
  \label{eq:tensor_decomposition}
\end{align}
We assume the external W and Z bosons satisfy the on-shell conditions, $\partial_\mu W^\mu = \partial_\mu Z^\mu = 0$.
These vertices come from the effective operators, e.g. $\mathrm{Tr}[\sigma_3 (D_\mu \mathbb{M}_1)^\dagger (D^\mu \mathbb{M}_2) ]$, which violate the custodial $\mathrm{SU}(2)_V$ symmetry.
In the 2HDM, these effective operators first appear at the one-loop level, especially if the scalar potential violates the custodial symmetry~\cite{Rizzo:1989ym,CapdequiPeyranere:1990qk,Kanemura:1997ej,Diaz-Cruz:2001thx,Arhrib:2006wd,Abbas:2018pfp,Aiko:2021can}.
Especially, the effects from $F_{\pm}$ are enhanced by the non-decoupling quantum effects of the additional scalar bosons~\cite{Kanemura:1997ej}.
In ref.~\cite{Kanemura:2024ezz}, we have shown all Feynman diagrams for the $H^\pm W^\mp Z$ vertices and the full formulae for $F_{\pm}$, $G_{\pm}$ and $H_{\pm}$ in the general 2HDM at the one-loop level.
As a result, we have found new contributions to these vertices from the imaginary part of the coupling constants.
In the following, we show the results from such contributions.

Through these vertices, the decays $H^\pm \to W^\pm Z$ are possible, if it is kinematically allowed.
The decay rates are given by 
\begin{align}
  \Gamma (H^\pm \to W^\pm Z) = \frac{m_{H^\pm} \lambda^{\frac{1}{2}}(1,w,z) }{16 \pi} \Big( |\mathcal{M}_{LL}|^2 +|\mathcal{M}_{TT}|^2  \Big),
\end{align}
where $w = m_W^2 / m_{H^\pm}^2$, $z = m_Z^2 / m_{H^\pm}^2$ and $\lambda (a,b,c) = (a-b-c)^2 - 4 bc$.
The amplitudes from the longitudinal modes ($\mathcal{M}_{LL}$) and the transverse modes ($\mathcal{M}_{TT}$) are given by 
\begin{align}
  &|\mathcal{M}_{LL}|^2 = \frac{g^2}{4 z} \bigg| (1-w-z)F_{\pm} + \frac{\lambda(1,w,z)}{2w} G_{\pm} \bigg|^2, \notag \\
  &|\mathcal{M}_{TT}|^2 = g^2 \bigg( 2w |F_{\pm}|^2 + \frac{\lambda(1,w,z)}{2w} |H_{\pm}|^2 \bigg).
\end{align}

\begin{figure}[t]
  \centering
  \includegraphics[width=1\linewidth]{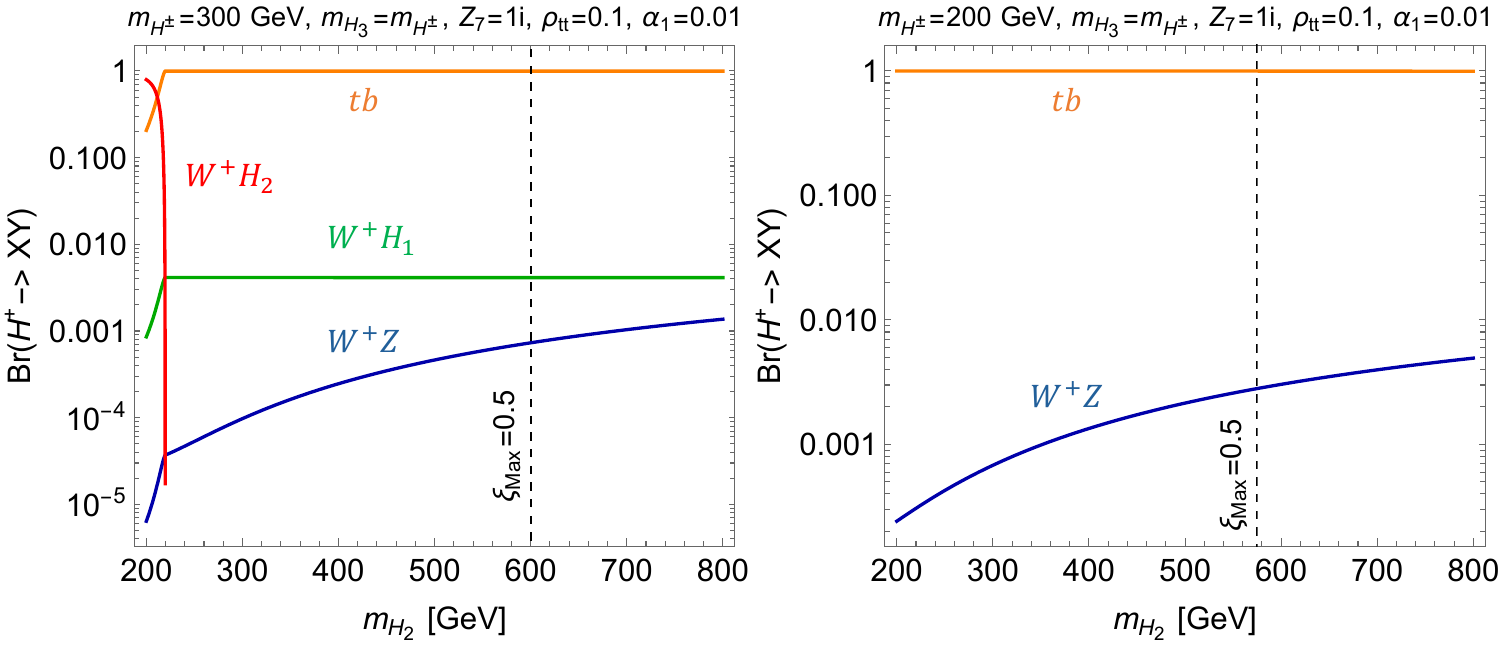}
  \caption{The branching ratios for $H^+ \to W^+Z$ (blue), $W^+ H_1$ (green), $W^+ H_2$ (red) and $tb$ (orange) as a function of $m_{H_2}$.
  The black dashed line shows the criterion, where the maximal $s$-wave scattering amplitude $\xi_{\mathrm{Max}}$ among the scalar and the longitudinal gauge bosons is 0.5.
  }
  \label{fig:decayrate}
\end{figure}
In fig.~\ref{fig:decayrate}, $\mathrm{Br}(H^+ \to XY)$ as a function of $m_{H_2}$ is shown.
Each of the decay modes $H^+ \to W^+ Z$, $W^+ H_1$, $W^+ H_2$ and $tb$ is shown by the blue, green, red, and orange lines, respectively.
The black dashed line shows the criterion, where the maximal $s$-wave scattering amplitude $\xi_{\mathrm{Max}}$ among the scalar and the longitudinal gauge bosons is 0.5~\cite{Lee:1977eg,Gunion:1989we,Kanemura:1993hm,Akeroyd:2000wc,Ginzburg:2005dt,Kanemura:2015ska}.
In the left (right) panel, $m_{H^\pm} = 300$~GeV ($m_{H^\pm} = 200$~GeV), $m_{H_3} = m_{H^\pm}$, $Z_7 = i$, $\rho_{tt} = 0.1$ and $\rho^f = 0$ except for $\rho_{tt}$ are taken.
The mixing angles $\alpha_1$ and $\alpha_2$ are taken by $0.01$ and $0$, respectively.
The violation of the custodial symmetry is related to the mass difference $m_{H_2} - m_{H^\pm}$, so that $\mathrm{Br}(H^+ \to W^+ Z)$ becomes large as growing $m_{H_2}$.
As shown in the left panel ($m_{H^\pm} = 300~\mathrm{GeV} > m_W + m_{H_1}$), due to non-zero $\alpha_1$, $\mathrm{Br}(H^+ \to W^+ Z)$ is suppressed by the decay $H^+ \to W^+ H_1$.
On the other hand, in the right panel where $m_{H^\pm} = 200~\mathrm{GeV}$ is taken, the decay $H^+ \to W^+ H_1$ is kinematically forbidden.
As a result, $\mathrm{Br}(H^+ \to W^+ Z) \gtrsim O(10^{-3})$ is realized for $\xi_{\mathrm{Max}} \le 0.5$.

In our model, the asymmetry between the decays $H^+ \to W^+ Z$ and $H^- \to W^- Z$ is caused by the interference of the scalar-loop and fermion-loop diagrams.
We define 
\begin{align}
  \Delta (H^\pm \to W^\pm Z) \equiv \Gamma(H^+ \to W^+ Z) - \Gamma(H^- \to W^- Z),
\end{align}
and the CP violating quantity~\cite{Arhrib:2007rm}
\begin{align}
  \delta_{\mathrm{CP}} \equiv \frac{\Gamma(H^+ \to W^+ Z) - \Gamma(H^- \to W^- Z)}{\Gamma(H^+ \to W^+ Z) + \Gamma(H^- \to W^- Z)}.
  \label{eq:deltaCP}
\end{align}
When $\rho^f = 0$ except for $\rho_{tt}$ and $Z_6 \ll 1$, each decay amplitude can be approximately written as 
\begin{align}
  &\mathcal{M}(H^+ \to W^+ Z) \simeq i \big(\rho_{tt}^R f_1 + Z_7^R (m_{H^\pm}^2 - m_{H_3}^2) f_2 \big) + \big( \rho_{tt}^I f_1 + Z_7^I  (m_{H^\pm}^2 - m_{H_2}^2) f_3 \big), \notag \\
  &\mathcal{M}(H^- \to W^- Z) \simeq -i \big(\rho_{tt}^R f_1 + Z_7^R (m_{H^\pm}^2 - m_{H_3}^2) f_2 \big) + \big(\rho_{tt}^I f_1 + Z_7^I (m_{H^\pm}^2 - m_{H_2}^2) f_3 \big),
\end{align}
where $f_{1,2,3}$ are mass dependent functions.
The fermion-loop functions in $f_1$ have the imaginary parts, so that those quantities can be expressed as  
\begin{align}
  &\delta_{\mathrm{CP}} \propto \Delta (H^\pm \to W^\pm Z) \propto |\mathcal{M}(H^+ \to W^+ Z)|^2 - |\mathcal{M}(H^- \to W^- Z)|^2 \notag \\
  &\propto \rho_{tt}^R Z_7^I (m_{H^\pm}^2 - m_{H_2}^2) f_3 \mathrm{Im}[f_1^*] + \rho_{tt}^I Z_7^R (m_{H^\pm}^2 - m_{H_3}^2) f_2 \mathrm{Im}[f_1].
\end{align}
Therefore, $\Delta (H^\pm \to W^\pm Z)$ and $\delta_{CP}$ are sensitive to the CP violating invariant $\mathrm{Im}[\rho_{tt} Z_7]$.
\begin{figure}[t]
  \centering
  \includegraphics[width=1.05\linewidth]{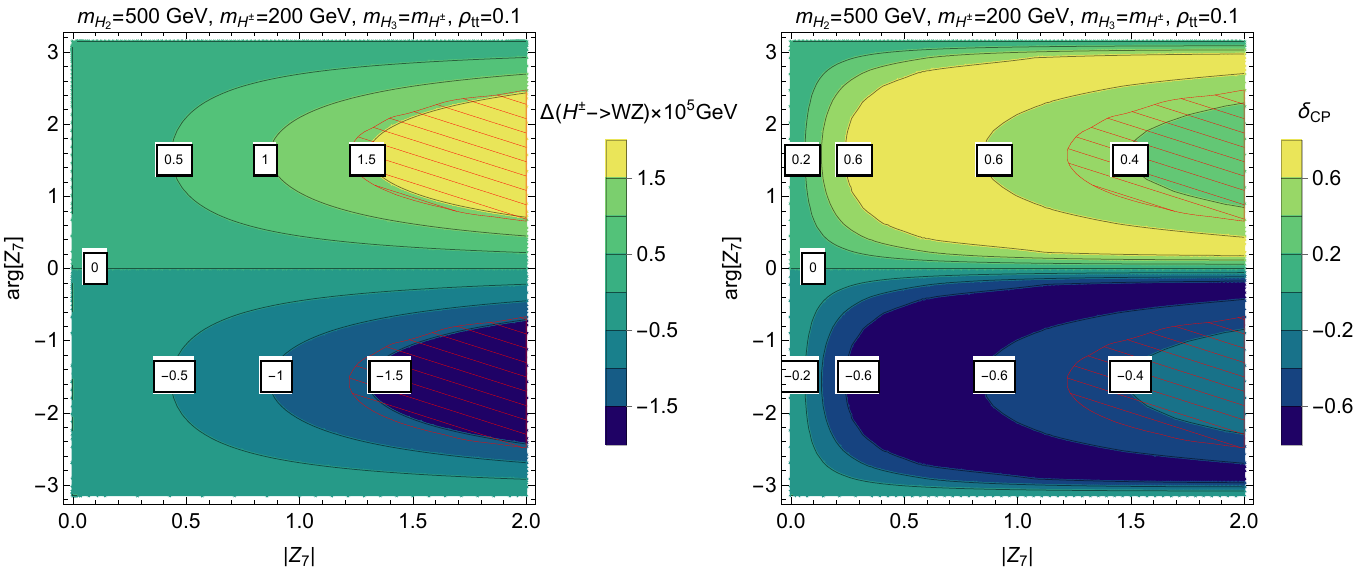}
  \caption{The contour plots for $\Delta (H^\pm \to W^\pm Z) \times 10^5$ GeV (left panel) and $\delta_{CP}$ (right panel) in $|Z_7|$-$\mathrm{arg}[Z_7]$ plane.
  The red shaded regions do not satisfy the bounded from the below condition.}
  \label{fig:CPV}
\end{figure}

In fig.~\ref{fig:CPV}, the contour figures of $\Delta (H^\pm \to W^\pm Z) \times 10^5$ GeV (left) and $\delta_{CP}$ (right) in the $|Z_7|$-$\mathrm{arg}[Z_7]$ plane are shown.
We set $m_{H^\pm} = 200$~GeV, $m_{H_2} = 500$~GeV, $m_{H_3} = m_{H^\pm}$, $\rho_{tt} = 0.1$, and $\alpha_1 = \alpha_2 = 0$.
The red shaded regions do not satisfy the bounded from the below condition~\cite{Klimenko:1984qx,Sher:1988mj,Nie:1998yn,Kanemura:1999xf,Ferreira:2004yd,Bahl:2022lio}.
At the points where the rephasing invariant $\mathrm{Im}[\rho_{tt} Z_7]$ takes the maximal (minimal) value, i.e. $\mathrm{arg}[\rho_{tt} Z_7] = \pi/2$ ($-\pi/2$), $\Delta (H^\pm \to W^\pm Z)$ and $\delta_{CP}$ take the maximal (minimal) value.
When $\delta_{CP} \simeq 0.6$, $\Gamma (H^- \to W^- Z) \simeq 1/4 \times \Gamma (H^+ \to W^+ Z)$ is shown by definition.

\section{Conclusion}
In these proceedings, we have summarized our results given in ref.~\cite{Kanemura:2024ezz} and discussed the loop induced $ H^\pm W^\mp Z$ vertices in the CP violating general two Higgs doublet model.
We have evaluated the CP violating effects to the decays $H^\pm \to W^\pm Z$ at the one-loop level.
We have found that the difference between the decays $H^+ \to W^+ Z$ and $H^- \to W^- Z$ is sensitive to the CP violating phases in the model.

\section*{Acknowledgments}
The work of S.~K. was supported by the JSPS Grants-in-Aid for Scientific Research No.~20H00160, No.~23K17691 and No.~24KF0060.
The work of Y.~M. was supported by the JSPS Grant-in-Aid for JSPS Fellows No.~23KJ1460.

\end{document}